\documentclass[preprint,12pt,authoryear]{elsarticle}

\usepackage{epsfig}

\usepackage{amssymb}

\usepackage{lineno}

\usepackage{xcolor}
\usepackage{amsmath}
\usepackage{chemformula}
\usepackage{setspace}
\journal{Earth and Planetary Science Letters}

\begin{document}

\begin{frontmatter}



\title{MgO Miscibility in Liquid Iron}


\author[label1,label2]{Leslie Insixiengmay}
\author[label1]{Lars Stixrude}

\affiliation[label1]{organization={Department of Earth, Planetary, and Space Sciences, University of California, Los Angeles},
            addressline={595 Charles E. Young Drive East}, 
            city={Los Angeles},
            postcode={90095}, 
            state={California},
            country={USA}}
\affiliation[label2]{organization={Now at Exponent Inc.},
            addressline={5401 McConnell Avenue}, 
            city={Los Angeles},
            postcode={90066}, 
            state={California},
            country={USA}}


\begin{abstract}

We explore phase equilbria on the MgO-Fe join as a prototype of lithophile-core interaction in terrestrial planets. Our simulations, based on density functional theory, are based on a two-phase method: fluids of initially pure MgO and Fe compositions are allowed to establish a dynamic equilbrium across a near-planar interface. Methods for analyzing the composition and other properties of the two coexisting phases show that MgO behaves as a component, with indistinguishable Mg and O concentrations in Fe-rich and oxide-rich phases. The phase diagram is well described as that of a symmetric regular solution, a picture confirmed by independent one-phase determinations of the enthalpy, entropy, and volume of mixing. The critical temperature, above which there is complete miscibility across the MgO-Fe join is 7000 K at 68 GPa, and 9000 K and 172 GPa. The rate of MgO exsolution from the Fe-rich liquid on cooling is similar to that found in previous experimental studies, and is too small to drive a dynamo.  

\end{abstract}

\begin{highlights}
\item The rate of MgO exsolution from liquid iron on cooling is too small to drive a dynamo.
\item The MgO-Fe system displays symmetric regular solution behavior.
\item Complete miscibility of MgO and iron occurs at temperatures greater than 7000 K at 60 GPa.
\item MgO dissolves as a component in liquid iron over the entire pressure and temperature range considered in this study without any cation exchange.
\end{highlights}


\end{frontmatter}


\section{Introduction}
\label{}

Lithophile elements have very limited solubility in Earth’s core today. However, the chemical interaction between core and mantle in the early Earth may have been more extensive (\citealp{chidester2022lithophile, Badro2016, Hirose2017}). Higher temperatures in the early Earth, due to accretional energy, and increased radioactive heat production, may have lead to much greater lithophile element solubility in the core. Lithophile elements may therefore serve as tracers of processes occurring during the hottest portions of Earth’s history. Super-Earths are expected to have much higher temperatures in their interiors than Earth and it is possible that lithophile element solubility in the Fe-rich cores of such bodies is more extensive than in Earth (\citealp{stixrude2014melting}). The temperature in the interiors of terrestrial planets early in their evolution may be sufficiently high that oxide and metals are completely miscible, forming a single homogeneous liquid (\citealp{walker1993superheating,Wahl2015}). Such a situation may have existed in portions of the earliest Earth, and may exist in the interiors of super-Earths.  


Hindering our knowledge of lithophile element solubility in metallic cores is the lack of a clear picture of the chemical reactions involved.  For example, it is not clear whether lithophiles should be viewed as dissolving as an oxide in the metal, dissociating as atoms in the metal, or exchanging with Fe \citep{Badro2018}.
These three pictures have contrasting implications for how we view the electronic structure and bonding of lithophile elements in the metal phase.

A better understanding of lithophile solubility in the core is also important for testing a model of the generation of Earth’s earliest magnetic field. Paleomagnetic evidence shows that the existence of Earth’s field extends back to at least 3.5 Ga \citep{biggin2009intensity} and possibly earlier \citep{Tarduno2015}, but how this field was produced prior to onset of inner core growth at $\sim$1.0 Ga \citep{labrosseetal_01,nimmo_15}, is unclear. It has been proposed that exsolution of lithophiles upon cooling of the core, and in particular Mg, may have driven the early dynamo (\citealp{Orourke2016, orourke2017thermal,Badro2018,wilson2022powering}). However, the rate of exsolution of lithophiles on cooling is still poorly constrained.
%

To address these issues, we explore the interaction between the most abundant lithophile component (MgO) and Fe metal over a wide range of pressure and temperature that encompasses their complete miscibility and the conditions expected during the lithophile exsolution that may have driven the early magnetic field. We perform two-phase simulations, which allow us to determine the way in which lithophile elements are incorporated in the metal, to quantify the composition of the two coexisting phases (Fe-rich and oxide-rich), their structure, and bonding.  To provide additional constraints on the phase diagram and to gain additional insight into core-mantle reaction, we also perform a series of homogeneous phase calculations to quantify the energetics of mixing, which confirm that the MgO-Fe system displays remarkably simple, symmetric regular solution behavior.  


     \section{Theory}



\begin{figure}[h!]
    \begin{center}
        \includegraphics[width=1.00\textwidth]{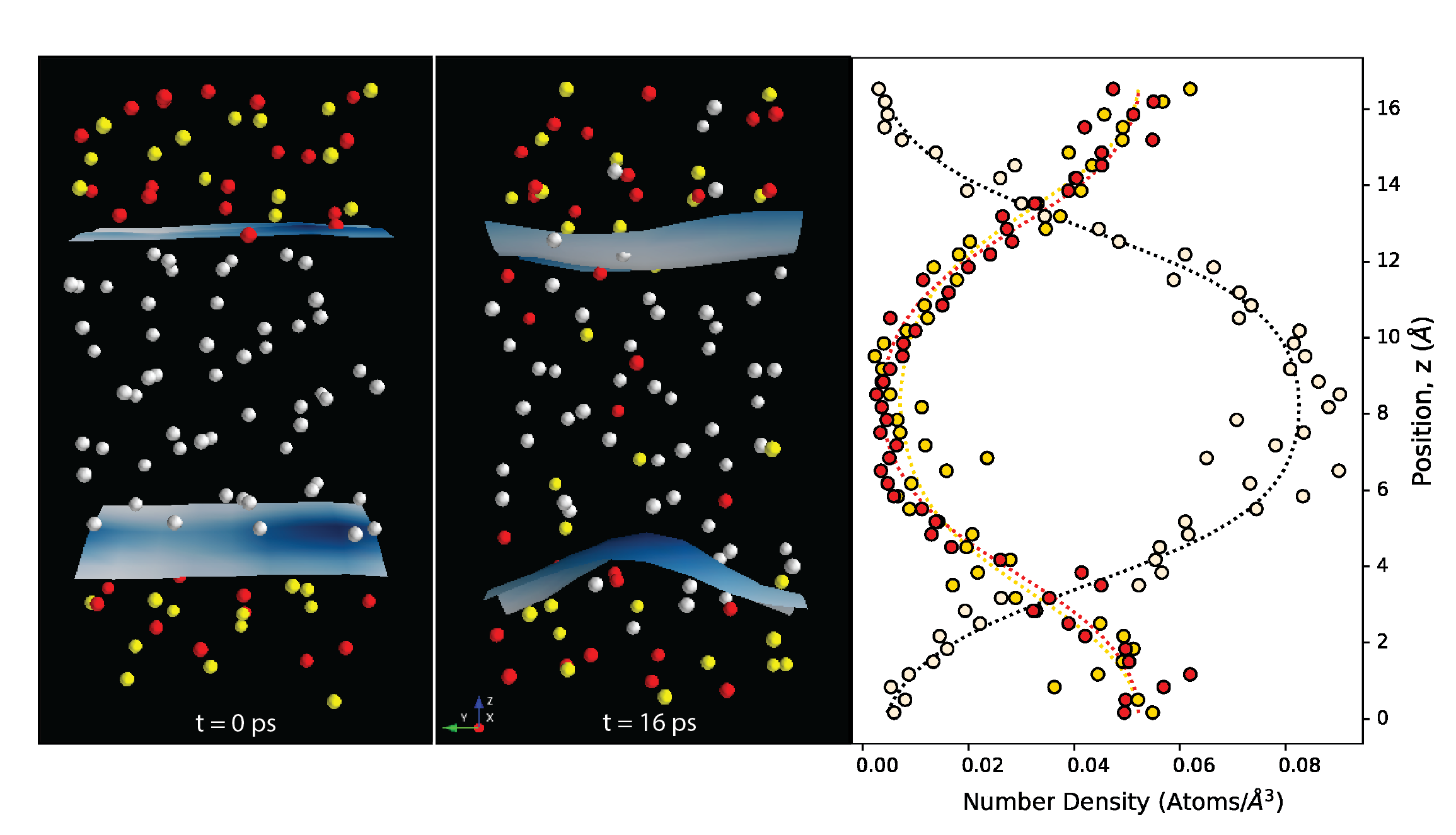}
        \caption{Simulation snapshots of the two-phase system at 6000 K and 60 GPa. Fe atoms are represented by tan spheres, Mg atoms by yellow spheres, and O by red spheres. The interface is illustrated by the blue surface, separating Fe-rich and oxide-rich phases. The initial configuration is on the left and an equilibrated snapshot at 16 ps in the center. The rightmost figure shows the one-dimensional density determined from the equilibrated portion of the simulation (16-18 ps) and the lines are fits to Eq.~\ref{eq_tanh}. }
        \label{intial_final_config}
    \end{center}
\end{figure}

Our two-phase simulations are initiated as domains of pure Fe liquid and pure MgO liquid joined at planar interfaces (Fig. \ref{intial_final_config} left).  During the course of the two phase simulations the system responds by establishing a dynamic equilibrium in which the composition of the two coexisting phases is stationary (Fig. \ref{intial_final_config} center; Supplementary Materials Fig. S2). We quantify the compositions of the two coexisting phases in two ways \citep{Xiao2018}. First, we use the one-dimensional density profile normal to the interface, which we find  follows the expected hyperbolic tangent form  (\citealp{Widom1982}):

\begin{equation}
    \rho (z) = \rho_2 + \frac{\rho_1 - \rho_2}{2} \sum_{j=1}^2 (-1)^j \tanh{\frac{(z-z_1)-\text{nint}(z-z_1) + (-1)^j w}{\delta}}
    \label{eq_tanh}
\end{equation}

\noindent where $\rho_1$ is the number density of an atom type (Fe, Mg, or O) in the phase whose center of mass is located at scaled coordinate $z_1$, $w$ is the half-width of the phase, $\rho_2$ is the number density of that atom type in the other phase, $\delta$ is the width of the interface, the $\text{nint}$ function accounts for periodic boundary conditions, and the sum accounts for the presence of two interfaces. An example of a fit of this form to our results is shown in Figure~\ref{intial_final_config} (right).

Second, we use the approach of \cite{Willard2010}. Define a coarse-grained density field by convolving the positions of Fe atoms with a Gaussian

 \begin{equation}
  \bar{\rho}(\textbf{r}, t) = \sum_i (2\pi\xi^2)^{-3/2} \exp \left[-\left(\mathbf{r}-\mathbf{r}_i \right)^2/2\xi\textsuperscript{2} \right] ,
  \label{eq_rhofield}
 \end{equation}

\noindent where $\mathbf{r}_i$ is the position of atom $i$, the sum is over the Fe atoms, and $\xi$ is the coarse-graining length, which we take to be 2.4 \r{A}. We locate the interface \textbf{s} at every time step such that 

\begin{equation}
    \bar{\rho}(\textbf{s},t) = c
\end{equation}

\noindent where we set the density of the interface $c$ to approximately 1/2 the bulk density of the Fe atoms. The interfaces determined in this way are shown in Figure~\ref{intial_final_config} (left, center). The interface remains quasi-planar throughout the course of the simulation, with the magnitude of fluctuations of the interface related to the surface tension (\citealp{buff1965interfacial}). The time-averaged interface is planar and represents the Gibbs dividing surface.     

With this definition of the interface, we can now assign each atom to one of the two phases. The proximity of each atom to the interface is

\begin{equation}
    a_i(t) = \{ \left[ \mathbf{s} - \mathbf{r}_i(t) \right] \cdot \mathbf{n}(t)\} |_{\mathbf{s}(t)=\mathbf{s}_i^*(t)}
    \label{eq_awc}
\end{equation}
where $\mathbf{s}_i^*(t)$ is the point on $\mathbf{s}(t)$ nearest to $\mathbf{r}_i(t)$ and $\mathbf{n}(t)$ is the unit vector normal to the interface in the direction of the local density gradient $\nabla \bar{\rho} (\mathbf{r},t) \ _{\mathbf{r}=\mathbf{s}^*(t)}$.  For definiteness and without loss of generality, we take the value of $a_i$ to be positive if atom $i$ is on the Fe-poor side of the interface and negative if on the Fe-rich side.  We can then count how many atoms of each type exist in each of the two phases at every time step of the simulation.  In performing this count, we exclude those atoms in the interfacial region, for which $| a_i | < \delta$.

We gain additional insight into our system by performing homogeneous molecular dynamics simulations at intermediate compositions on the MgO-Fe join. From these simulations, we determine the enthalpy $H$, the volume $V$, and the entropy $S$.  We find that our results are well represented by a symmetric regular solution, for which the Gibbs free energy of solution \citep{hill1986statmech}
\begin{equation}
    G(P,T,x) = W(P,T)x(1-x) + RT \left[ x \ln x + (1-x) \ln (1 - x) \right]
\end{equation}
and we assume the interaction parameter depends on pressure and temperature
\begin{equation}
    W(P,T) = W_H - T W_S + P W_V 
    \label{W_eq}
\end{equation}
so that the excess enthalpy, entropy, and volume of solution are, respectively
$\Delta H_{ex} = W_Hx(1-x)$, $\Delta S_{ex} = W_Sx(1-x)$, and $\Delta V_{ex} = W_Vx(1-x)$.
$R$ is the universal gas constant and $x$ is the composition, which we take to be the mole fraction of MgO $x=[\text{MgO}]/([\text{MgO}] + [\text{Fe}])$.  

In equilibrium, the compositions of the two coexisting phases: $x$ and $(1-x)$ can be derived from the root $\partial G / \partial x$ = 0
    \begin{equation}
       \frac{x}{1-x} = \exp \left[ -\frac{W(P,T)(1-2x)}{RT} \right]
        \label{solvus_eq}
    \end{equation}
The critical temperature $T_C$ above which a single phase exists, occurs at $x=1-x=1/2$.  Combining Eqs.~\ref{W_eq},\ref{solvus_eq} 
\begin{equation}
    T_C(P) = \frac{W_H + PW_V}{2R + W_S}
\end{equation}
so that $dT_C / dP = W_V/(2R + W_S)$. We find the rate of exsolution of MgO from the metal phase by taking the implicit temperature derivative of  \ref{solvus_eq} and solving for
\begin{equation}
\frac{\partial x}{\partial T} = \frac{1 - 2 x}{R T^2} (W + W_ST) \left[ \frac{1}{x(1-x)} - \frac{2W}{RT}\right]^{-1}
\label{exsolve_eq}
\end{equation}
or in the limit $x \rightarrow 0$
\begin{equation}
    \frac{\partial \ln x}{\partial T} = \frac{W + W_ST}{RT^2}
\end{equation}
to first order in $x$.
In the geophysical context one typically expresses the exsolution rate in terms of mass fraction $\chi=\left[ 1 + m(1-x)/x \right]^{-1}$ rather than mole fraction, so that the mass exsolution rate (units: g/g/K or K$^{-1}$)
\begin{equation}
    C=\frac{\partial \chi}{\partial T} = m \frac{\chi^2}{x^2}\frac{\partial x}{\partial T}
\end{equation}
Here $m=M_{Fe}/M_{MgO}$ is the ratio of the molar masses of the two components.  In the limit of small $x$, $\partial \ln \chi / \partial T = \partial \ln x / \partial T$.

While our focus is on liquid-liquid phase equilibria, and all of our simulations are in the liquid state, for completeness, we can also use our results to estimate solid-liquid equilibria as well. The MgO content of liquid coexisting with (pure) MgO crystal is governed by

    \begin{equation}
        W(1-2x) + RT \ln x + G_{liq}^{MgO} = G_{xtl}^{MgO}
        \label{liq_sol_MgO}
    \end{equation}

\noindent where $G_{liq}^{MgO}$ and $G_{sol}^{MgO}$ are the Gibbs free energy of pure liquid and solid MgO, respectively, and $G_{xtl}^{MgO} -G_{liq}^{MgO}= \Delta S_m (T - T_m)$. For the purpose of this calculation, we adopt values of the temperature $T_m$ and entropy $\Delta S_m$ of melting of pure MgO from the study of \cite{alfe2005melting}.

We also use one phase simulations to derive additional insight into the structure of oxide-metal fluids on the MgO-Fe join by computing the radial distribution function \citep{mcquarrie1975statistical}, ionic charges, and bond lifetimes. We compute ionic charges according to the Bader analysis \citep{tang2009grid, sanville2007improved, henkelman2006fast, yu2011accurate}. The lifetime of each atom-pair is estimated from 

    \begin{equation}
       \tau = \int_0^\infty \beta (t)dt'.
        \label{bac_life}
    \end{equation}

\noindent with bond auto-correlation function

\begin{linenomath}
    \begin{equation}
       \beta(t) = \bigg \langle \frac{b_{ij}(t_0) \cdot b_{ij}(t_0 + t) }{b_{ij}(t_0)^2} \bigg \rangle 
        \label{bacf}
    \end{equation}
\end{linenomath}

\noindent where $b_{ij}(t) = 1$ if a bond between atoms $i$ and $j$ exist at time $t$, and $b_{ij}(t) = 0$ otherwise.  We also compute the histogram of bond lifetimes.  We take the bond cutoff criteria to be the distance to the first minimum in the corresponding radial distribution function.  

     \section{Computation}

All simulations are based on density functional theory, using the projector augmented wave (PAW) method (\citealp{Kresse1999}) as implemented in the Vienna \textit{ab initio} Simulation Package (VASP) (\citealp{Kresse1996}). We use the PBEsol generalized gradient approximation (\citealp{Perdew2008a}), which we have previously shown to yield excellent agreement with experiment in Fe-bearing oxides (\citealp{Holmstrom2015}). To account for strong correlation, we use the +U method (\citealp{Anisimov1997}) with $U-J=$2.5 eV (\citealp{Holmstrom2015}). We use PAW potentials of 14, 8, and 6 valence electrons of Fe, Mg, and O with core radii of 1.16, 1.06, and 0.82 \AA, respectively. Our simulations are spin-polarized.  In order to compare with previous results, which were non-spin polarized, we also perform non-spin polarized simulations. We sample the Brillouin zone at the Gamma point and use a plane-wave cutoff of 500 eV, which we find yields pressure and energy convergence to within 0.2 GPa and 5 meV/atom, respectively (Supplementary Material Fig. S1). 
We assume thermal equilibrium between ions and electrons via the Mermin functional (\citealp{mermin1965thermal}), and thermodynamic averages are computed after discarding 20 $\%$ of the time steps to allow for transients; uncertainties are computed via the blocking method (\citealp{Flyvbjerg1989}).

We perform two-phase simulations in the canonical ensemble (constant $NVT$) using a Nóse-Hoover thermostat (\citealp{nose1984unified,nose1991constant,hoover1985canonical,frenkel2001understanding}) with a time step of 1 fs for a duration of 10-25 ps. 
We prepare the initial condition by first performing a homogeneous simulation of MgO liquid at temperature $T$ in a cubic simulation cell of linear dimension $L$ and 31 MgO units.  We then perform a homogeneous simulation of Fe liquid at $T$ in a simulation cell of identical dimensions and adjust the number of Fe atoms until the pressure matches that of the MgO simulation.  Finally, we form the initial condition of the two-phase simulation by combining equilibrated snapshots of the MgO and Fe simulations, forming a similation cell of dimension $L \times L \times 2L$ (Fig. 1 left).
In the lower pressure regime, the volume of the cell is 1162.83 \AA$^3$  and contains 31 MgO units and 55 Fe atoms, while in the higher pressure regime, the volume of the cell is 910.12 \AA$^3$ and has the same numbers of atoms.  
In these isochoric simulations the pressure increases linearly on heating: the pressure of equilibrated systems range from 55 GPa (5000 K) to 68 GPa (7000 K) in the lower pressure regime and from 136 GPa (5000 K) to 172 GPa (9000 K) in the higher pressure regime (Supplementary Material Table S1).  

For the homogeneous systems our simulations are in the $NPT$ ensemble (\citealp{parrinello1980crystal,parrinello1981polymorphic}), with a Langevin thermostat (\citealp{allen2017computer,hoover1982high,evans1983computer}), and a time step of 1 fs for 15 ps. These simulations were run at 60 GPa and 10,000 K and consisted of 64 atoms comprising several compositions across the MgO-Fe join. We determine the enthalpy and the volume from these $NPT$ simulations. For the entropy, we use the two-phase thermodynamic - memory function (2PT-MF) method , which decomposes the vibrational density of states into solid-like and fluid-like portions, using a memory function-based formalism \citep{desjarlais2013first,Wilson2021a}.  For determination of the entropy, we found that the $NPT$ simulations were not appropriate as the Langevin thermostat biases the vibrational density of states. Therefore, once we determine the equilibrium volume at a given pressure, we continue at that volume in the canonical ensemble for an additional 15 ps, from which we derive the absolute entropy using the 2PT-MF method.  Homogeneous simulation results are summarized in Supplementary Materials Table S3.


\section{Results}

We find two coexisting phases in dynamic equilibrium each with a stationary composition: one Fe-rich and one oxide-rich. In both phases, the concentration of Mg and O are equal to each other, i.e. MgO behaves as a component at all pressure-temperature conditions we investigated. Figure \ref{intial_final_config} shows the result of a typical simulation at 60 GPa and 6000 K, showing small, non-zero concentration of MgO in the Fe-rich phase and a nearly equal concentration of Fe in the MgO-rich phase.   

\begin{figure}
    \begin{center}
        \includegraphics[width=0.8\textwidth]{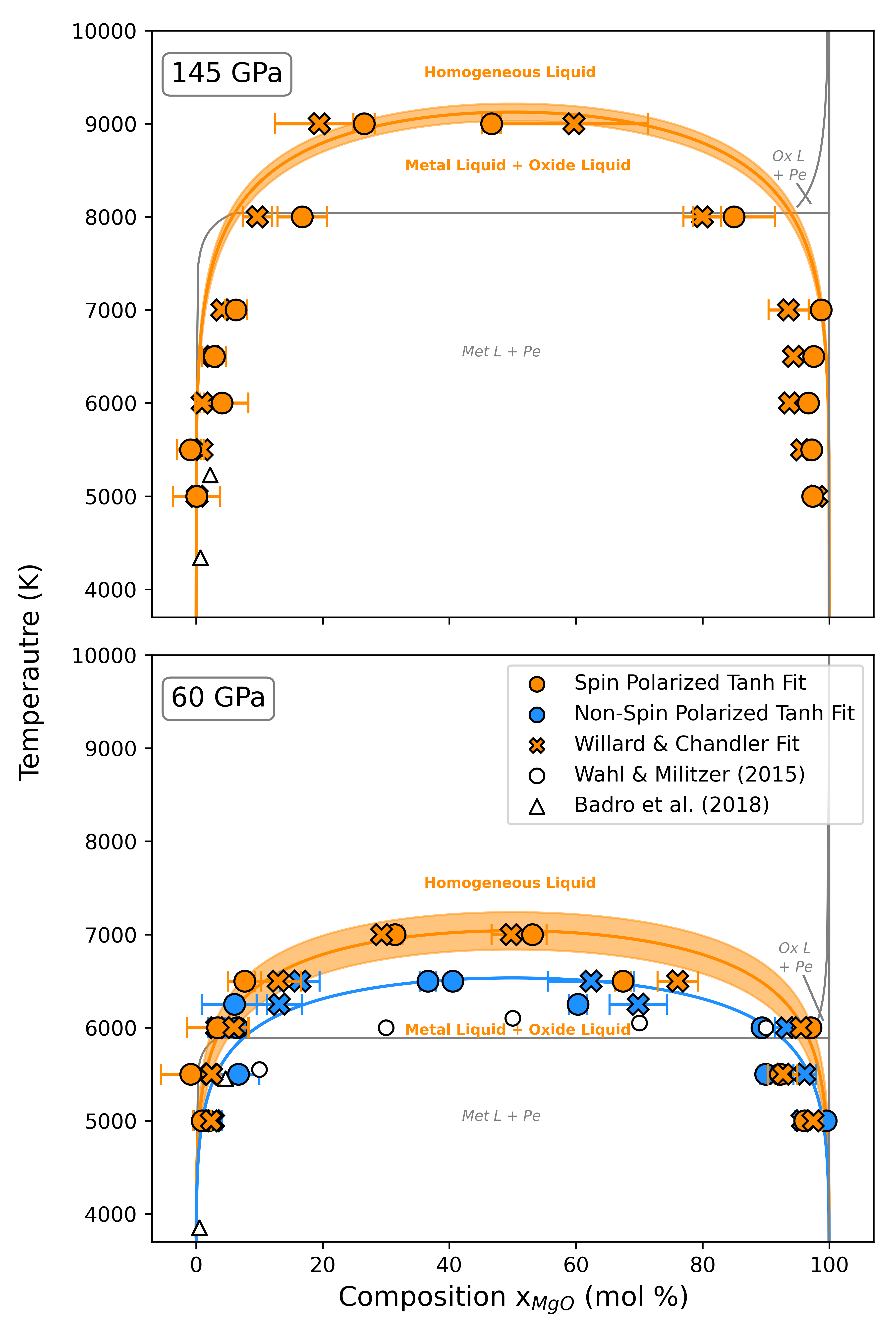}
        \caption{Phase diagrams in the low (bottom) and high (top) pressure regimes. Composition determined from Eq.~\ref{eq_tanh} are in circles (orange = spin-polarized and blue = non-spin polarized), and from the \cite{Willard2010} method in X's. White symbols represent previous studies: theoretical results from \cite{Wahl2015} in circles and experimental data from \cite{Badro2018} in triangles. The orange curve is the regular symmetric solvus computed from equation \ref{solvus_eq}, using values of $W_H$, $W_S$ and $W_V$ determined from our homogeneous simulations.  The gray lines represent the computed phase diagram including the stability of crystalline MgO.  Orange and gray curves account for the increase of pressure with increasing temperature.  The figures are labelled by the pressure at 6000 K.}
        \label{stacked_phase_diagram}
    \end{center}
\end{figure}

The concentration of MgO in the Fe-rich phase and the concentration of Fe in the MgO-rich phase increase on heating (Fig. \ref{stacked_phase_diagram}). The two phases become completely miscible at a critical temperature of 7000 K at 68 GPa, and at 9000 K at 172 GPa. Our results are consistent with symmetric regular solution behavior. 
We find good agreement with previous experiments \citep{Badro2018}, but disagree with a previous theoretical study, which found a significantly lower critical temperature \citep{Wahl2015}. We attribute the latter disagreement to the neglect of spin-polarization in the previous study. To test the significance of spin-polarization, we also determined the phase diagram with non spin-polarized simulations (Supplementary Materials Table S2) and find a significantly lower critical temperature, in better agreement with \cite{Wahl2015} (Fig. \ref{stacked_phase_diagram}).     

\begin{figure}[h!]
    \begin{center}
        \includegraphics[width=1\textwidth]{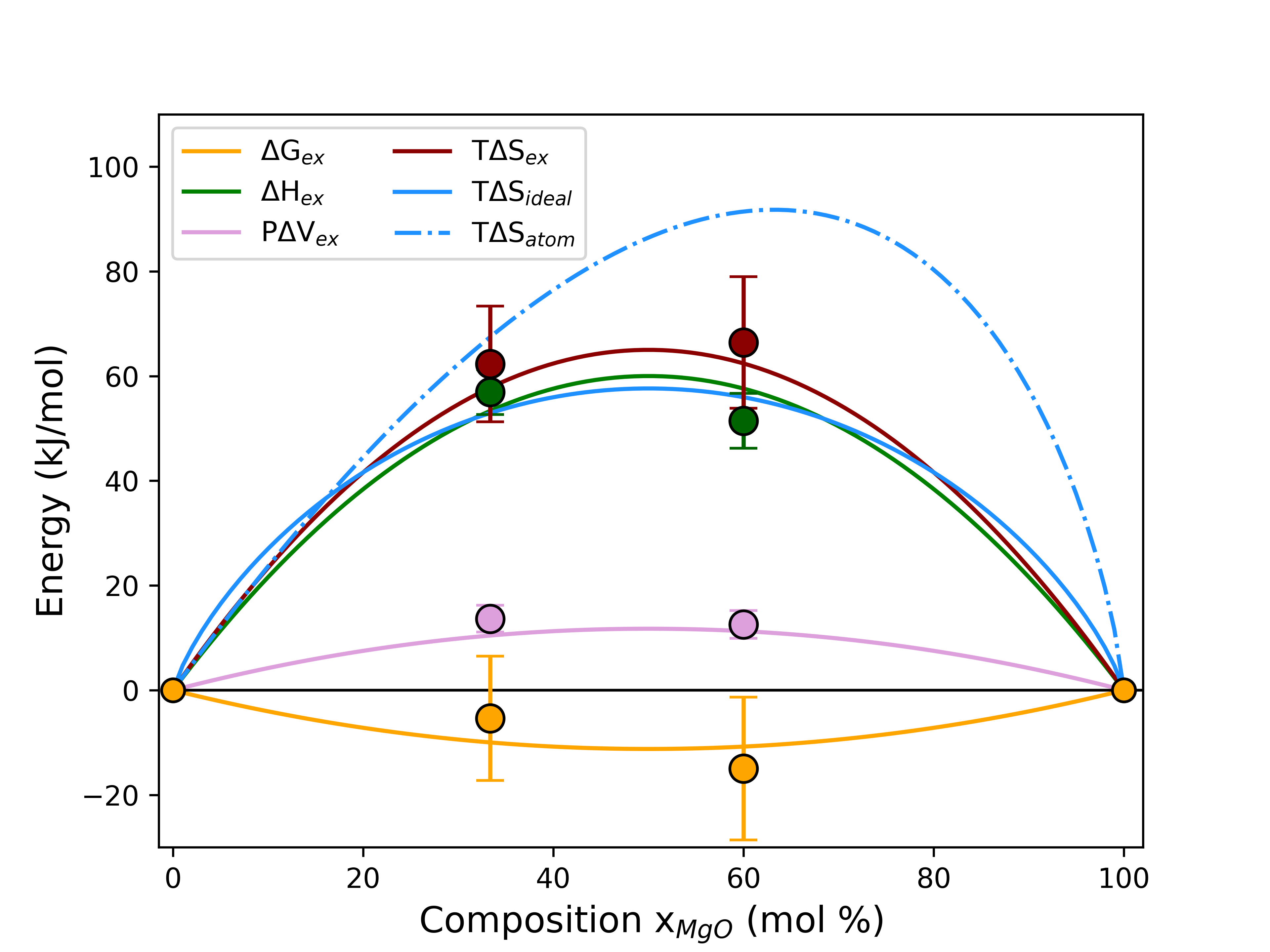}
        \caption{Energetics of mixing at 60 GPa and 10,000 K, showing the results of our simulations (symbols) and symmetric regular solution fits (lines) for the enthalpy (green), the entropy (red), and the volume (purple).  We also compute the excess Gibbs free energy of mixing $\Delta G_{ex}=\Delta H_{ex} -T \Delta S_{ex} + P V_{ex}$ (orange).  The symmetric regular solution fits yield $W_H$ = 240 $\pm$ 12.01 kJ/mol, $W_S$ = 26 $\pm$ 0.10 J/(K $\cdot$ mol), and $W_V$ = 1.3 $\pm$ 0.07 \AA$^3$/mol. We compare with the ideal entropy of mixing assuming Fe and MgO as components (blue solid line) and assuming Fe, Mg, and O as components (blue dashed line).}
        \label{mix_params}
    \end{center}
\end{figure}

We compute the phase diagram independently, from our determinations of the energetics of mixing in homogeneous systems, and find excellent agreement with our two-phase simulations (Fig. \ref{stacked_phase_diagram}). The energetics of mixing show that immiscibility originates in a positive enthalpy of mixing, and that the increase of the critical temperature with increasing pressure originates in a positive volume of mixing (Fig. \ref{mix_params}). The excess entropy of mixing is also positive and similar in magnitude to the ideal entropy of mixing of MgO and Fe components (Fig. \ref{mix_params}). We confirm that the Gibbs free energy of mixing is negative across the join at super-critical temperature (Fig. \ref{mix_params}), consistent with our finding of a single homogeneous fluid in our two-phase simulations.       

\begin{figure}[h!]
    \begin{center}
        \includegraphics[width=1\textwidth]{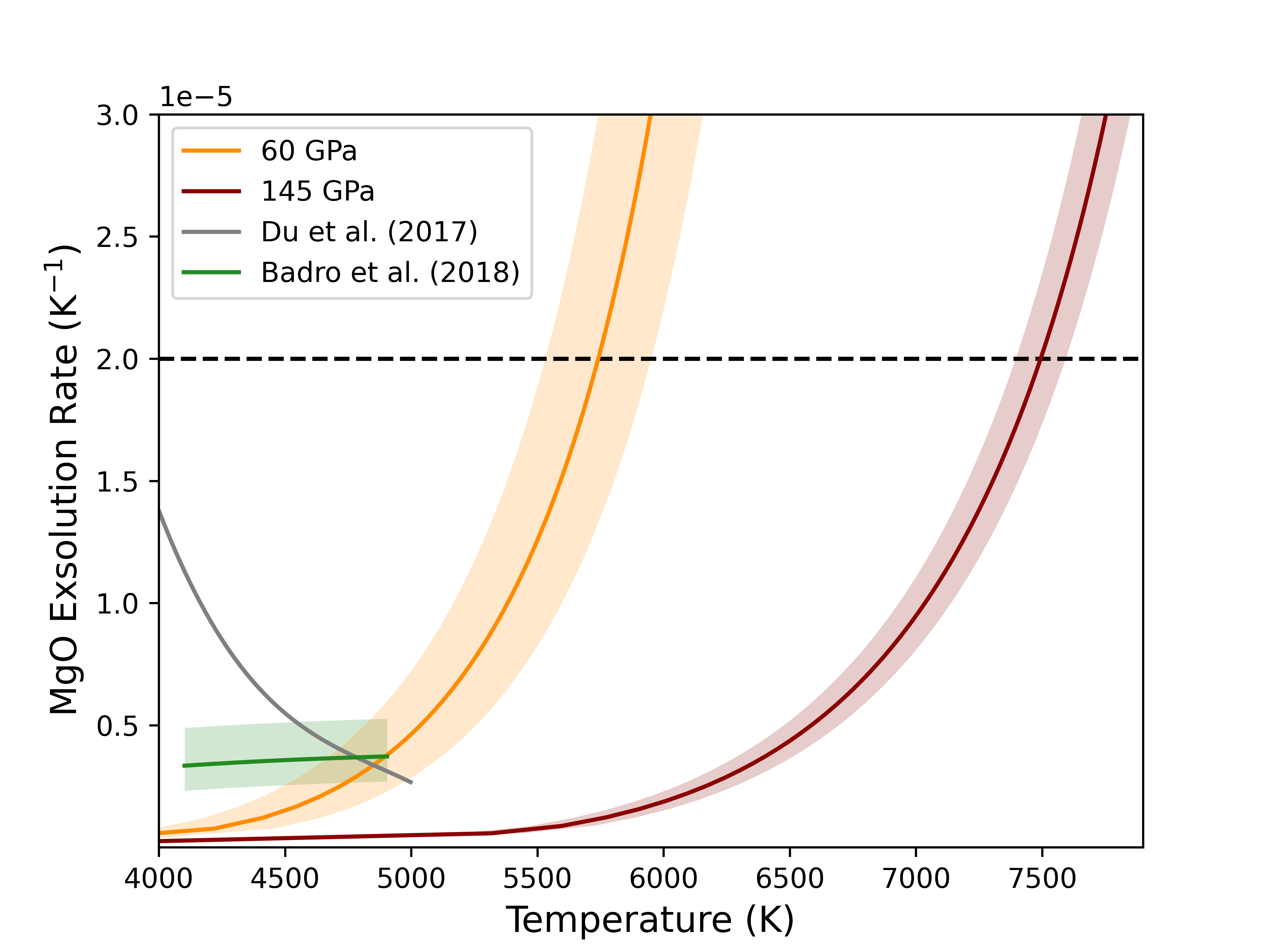}
        \caption{Exsolution rate of MgO from the Fe-rich fluid at 60 GPa (orange) and 145 GPa (red). Exsolution rates are compared with previous experimental studies from \cite{Du2017} (gray) and \cite{Badro2018} (green). The dashed line represents the smallest exsolution rate that produced a dynamo in the study of \cite{orourke2017thermal}.}
        \label{exsolution_rate}
    \end{center}
\end{figure}

The exsolution rate increases with increasing temperature and decreases with increasing pressure (Fig.  \ref{exsolution_rate}). The exsolution rate is less than $2$ x $10^{-5}$ K$^{-1}$ at all temperatures less than 5500 K at 60 GPa and less than 7500 K at 145 GPa. Our results at 60 GPa are similar in magnitude to previous experimental results \citep{Du2017,Badro2018}.  The study of \cite{Badro2018} also find that the exsolution rate increases with increasing temperature, albeit more gradually than what we find, while the study of \cite{Du2017} find that the exsolution rate decreases with increasing temperature.

\begin{figure}[h!]
    \begin{center}
        \includegraphics[width=1\textwidth]{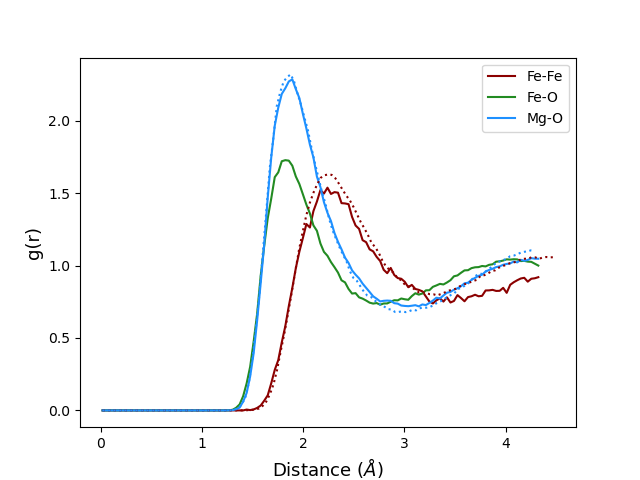}
        \caption{Radial distribution functions of the homogeneous fluid with $x_{MgO}$=0.6 at 10000 K and 60 GPa (solid lines) compared with the same pair distributions in the pure phases (dashed lines).}
        \label{rdf}
    \end{center}
\end{figure}

The radial distribution function gives us additional insight into the microscopic interactions between components (Fig. \ref{rdf}). The radial distribution function of the mixed homogeneous fluid ($x$=0.6) at 10,000 K and 60 GPa, shows distinct peaks for Mg-O, Fe-O, and Fe-Fe interactions. We compare these with the radial distribution functions computed for pure Fe, and pure MgO fluids at the same conditions. We find that $g_{Mg-O}$ and $g_{Fe-Fe}$ are very similar in the mixed and the pure fluids: in the mixed fluid, the first coordination shell is slightly less distinct: the height of the first maximum in $g(r)$ is slightly less and the first minimum occurs at larger distances. The Mg-O coordination number in the mixed fluid is much greater than that of the Fe-O coordination number (4.2 vs. 3.1)

We find that bonds within the first coordination shell have very short lifetimes: for example, in the homogeneous $x$=0.6 fluid, the mean lifetime of Mg-O bonds is 155 fs, as compared with 60 fs for the vibrational period of the TO mode in periclase at a similar pressure (Supplementary Materials Fig. S3).  

\begin{figure}[h!]
    \begin{center}
        \includegraphics[width=1\textwidth]{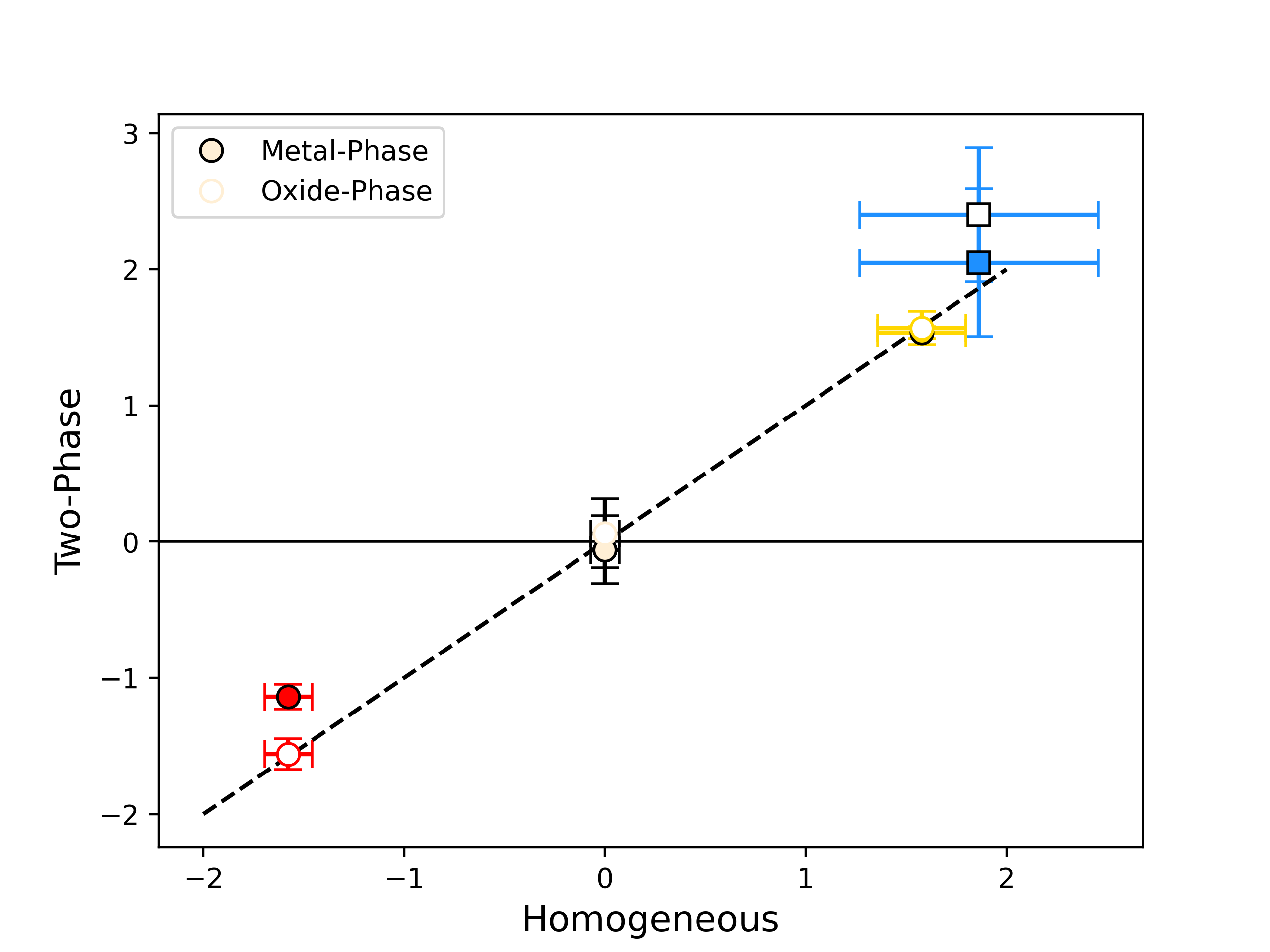}
        \caption{Mean values of the ionic charge for each atom type (Mg: yellow, O: red, Fe: tan) and the magnetic moment of Fe (blue squares) in each of the two phases (closed: Fe-rich phase, open: MgO-rich phase) from our two-phase simulation at 60 GPa, 6000 K, compared with their values in the corresponding pure phases (pure MgO or pure Fe liquid) at the same conditions. Error bars indicate the standard deviation in the value. The one:one line is shown as dashed black.  At 8000 K and 162 GPa, the absolute magnetic moments for the metal and oxide phases are 0.36 $\pm$ 0.47 and 0.68 $\pm$ 0.65 $\mu_B$, respectively.}
        \label{measured_expected}
    \end{center}
\end{figure}

Ionic charges are similar in oxide-rich and Fe-rich phases (Fig. \ref{measured_expected}).  The ionic charge of Mg and Fe are the same within standard deviation in MgO-rich and Fe-rich phases, while in the case of oxygen, the charge is slightly less negative in the Fe-rich phase (by 0.3).  Moreover, the ionic charges of the atoms are very similar to their values in the corresponding pure phases (pure MgO or pure Fe).  The magnetic moment of Fe atoms is somewhat larger in the oxide phase as compared with the metal phase (by 0.2 Bohr magnetons).

\section{Discussion}

Previous studies leave uncertain the nature of the interaction between MgO and Fe, as epitomized by the chemical reactions \citep{Badro2018}


    \begin{equation}
       \textrm{Dissolution: } \ch{MgO^{oxide}} \rightleftharpoons \ch{MgO^{metal}}
        \label{dissolution}
    \end{equation}
    
    \begin{equation}
       \textrm{Dissociation: } \ch{MgO^{oxide}} \rightleftharpoons \ch{Mg^{metal}} + \ch{O^{metal}}
        \label{dissociation}
    \end{equation}

    \begin{equation}
       \textrm{Exchange: } \ch{MgO^{oxide}} + \ch{Fe^{metal}} \rightleftharpoons \ch{FeO^{oxide}} + \ch{Mg^{metal}}
        \label{exchange}
    \end{equation}

\noindent The first views MgO as dissolving as a component in the metal phase.  The second also involves dissolution, but views Mg and O as separate, dissociated components once dissolved in the metal. The third has also been widely used to understand the partitioning of moderately siderophile elements between oxide and metal (\citealp{wood2008core, wood2008accretion}). Previous experimental studies disagree on whether the exchange reaction, or the other two, can best represent experimental measurements (\citealp{Du2017,Badro2018,Chidester2022}). Previous theoretical studies have assumed that MgO is incorporated in Fe-rich liquid either via dissolution or dissociation, not via exchange \citep{Wahl2015,wilson2022powering}.

Our results point towards the dissolution mechanism as the best representation of MgO-Fe reaction. We find that MgO behaves as a component, with Mg and O concentrations equal to each other in both oxide-rich and Fe-rich phases over the entire range of pressure and temperature that we have considered, excluding the exchange mechanism. Our results favor the dissolution mechanism over the dissociation mechanism. In the dissociation mechanism, Mg and O are viewed as separate components, with implications for the ideal entropy of mixing. Whereas we find that the phase diagram is symmetric, consistent with mixing between MgO and Fe components and the dissolution mechanism, ideal mixing among Mg, O, and Fe, introduces asymmetry to the ideal entropy of mixing on the MgO-Fe join that is inconsistent with our results (Fig. \ref{mix_params}).  

The dissolution mechanism does not imply covalent bonding between Mg and O, as argued by \cite{Badro2018}. Indeed, we find no such bonding, with ionic Mg-O bonds surviving for no more than two vibrational periods on average, and O atoms rapidly exchanging within the first coordination shell of the Mg cation (Supplementary Materials Fig. S3). Moreover, the ionic charges of Mg and O are very similar in the metal and the oxide phases, indicating that Mg and O are still ionically, rather than covalently bonded in the metal phase.

We find that the rate of MgO exsolution from the metal phase is not sufficient to drive the early dynamo.  The smallest exsolution rate that produced a dynamo in the study of \cite{orourke2017thermal} is $C=2 \times 10^{-5}$ K$^{-1}$.  However, we find that $C$ only exceeds this value for core-mantle boundary temperatures $>$7500 K (Fig.~\ref{exsolution_rate}).  Temperatures this high are not typically considered in thermal evolution models, and while not impossible during or very shortly after Earth's main accretion phase, could not have been sustained for long.  It is worth pointing out that other important uncertainties remain in our evaluation of the thermal evolution of the core including the thermal conductivity of the outer core, which may affect the minimum value of $C$ required to drive a dynamo.

We have investigated a simplified chemical system and it is possible that the rate of exsolution of MgO in a multi-component system may differ from what we find. However, experiments show that the activity of Mg in the metal phase depends little on the presence of other elements. For example \cite{Chidester2022} find that the activity of Mg in the metal phase is independent of the concentration of other elements except for O, for which they find a strong affinity for Mg (as expressed in their formulation by a negative value of the interaction parameter $\epsilon^O_{Mg}$), in accord with our findings that Mg and O are strongly associated.  On the other hand, the same study found that the activity of MgO in the oxide phase does depend on the silica concentration in the oxide phase, with a value of the interaction parameter $W_{MgO-SiO_2}$ that is in excellent agreement with previous theoretical studies \citep{dekoker2013thermodynamics}. It is also possible that the exsolution of other components, such as SiO$_2$, may be sufficient to drive an early dynamo \citep{Hirose2017}. 


 An important issue related to exsolution of light elements that has received relatively little attention is the pressure dependence of exsolution. The exsolution-driven dynamo scenario envisions exsolution first occurring at the top of the core: exsolution of lithophiles produces denser Fe-enriched residual fluid, which then drives large scale fluid motion.  Exsolution beginning at the top of the core is in accord with experimental findings that the pressure-dependence of exsolution is either weak or undetectable \citep{Orourke2016,Badro2018,Chidester2022}.  However, we find significant pressure dependence:  $dT_C/dP=24$ K/GPa, which is much greater than the expected adiabatic gradient in the core $\Gamma=7$ K/GPa. 
 If $dT_c/dP>\Gamma$ as we find, then exsolution begins at the bottom of the core.  It has been argued that in this scenario, the exsolution-driven dynamo does not operate since the Fe-enriched residual fluid, which is first produced at the center of the core, does not drive large-scale fluid motion or produce a dynamo \citep{landeau2022sustaining}.  Further dynamical studies should be undertaken to explore the effects of a strong pressure dependence of lithophile exsolution.  The origin of the apparent discrepancy between our results and experiment is not clear. It is possible that experimental detection of pressure dependence is confounded by co-variance with other experimental parameters together with pressure, such as temperature, and bulk composition. It is also possible that the pressure dependence of exsolution is diminished in systems that are more chemically rich than the simplified system that we have focused on.

\section{Conclusions}

Lithophile element solubility in the core may have been much more extensive in the early Earth when temperatures were hotter, or in Super-Earths, where the larger planetary size may lead to much higher interior temperatures. The critical temperature that we find (7000 K at 68 GPa) means that complete miscibility of Fe and MgO cannot be excluded very early in Earth's thermal evolution. By examining the entire MgO-Fe join in the liquid state, we find remarkably simple behavior: a symmetric regular solution between the two end-members.  Our simulations favor the dissolution picture of lithophile-Fe interaction, in which MgO is viewed as a component that is not dissociated. Mg and O remain ions of near nominal charge in the metal liquid and Mg-O bonds last for only a short time.  

Our results do not support the notion that the early magnetic field was generated by exsolution of MgO from the core. The rate of exsolution on cooling that we find is too small. Moreover, we find that exsolution likely initiated at the bottom of the core, rather than the top, in which case exsolution would not have provided a driving force for fluid motions. Other mechanisms for producing the Earth's earliest field should be further explored, including the exsolution of lithophiles other than Mg from the core, or a silicate dynamo \citep{Stixrude2020}.



\section{Acknowledgements and Funding}

This project is supported by the National Science Foundation under grant EAR-2223935 to LS. Calculations were carried out using the Hoffman2 Shared Cluster provided by UCLA Institute for Digital Research and Education's Research Technology Group.




\newpage

\bibliographystyle{elsarticle-harv} 
\bibliography{references}





\end{document}


\begin{frontmatter}



\title{Supplementary Material for: MgO Miscibility in Liquid Iron}


\author[label1]{Leslie Insixiengmay}
\author[label1]{Lars Stixrude}

\affiliation[label1]{organization={Department of Earth, Planetary, and Space Sciences, University of California, Los Angeles},
            addressline={595 Charles E. Young Drive East}, 
            city={Los Angeles},
            postcode={90095}, 
            state={California},
            country={USA}}


\end{frontmatter}



\renewcommand{\thetable}{S\arabic{table}}  
\renewcommand{\thefigure}{S\arabic{figure}}
\renewcommand{\thesection}{S\arabic{section}}

\section*{Two-phase simulation results}

\begin{table}[h!]
    \centering
    \begin{tabular}{c c c c }
        \hline
        \hline
        Pressure (GPa) & Volume (\AA$^{3}$) & Temperature (K) & Internal Energy (eV) \\
        \hline
        55.049 $\pm$ 0.51 & 1162.83 & 5000  & -584.546 $\pm$ 3.860 \\
        59.166 $\pm$ 0.15 & $\cdot$ & 5500  & -561.600 $\pm$ 0.741 \\
        61.321 $\pm$ 0.21 & $\cdot$ & 6000  & -546.705 $\pm$ 1.935 \\
        66.043 $\pm$ 0.27 & $\cdot$ & 6500  & -511.744 $\pm$ 2.298 \\
        68.465 $\pm$ 0.40 & $\cdot$ & 7000  & -494.073 $\pm$ 3.297 \\
        \hline
        135.585 $\pm$ 0.42 & 910.12 & 5000  & -495.586 $\pm$ 1.981 \\
        140.471 $\pm$ 0.84 & $\cdot$ & 5500  & -475.381 $\pm$ 3.936 \\
        144.755 $\pm$ 0.22 & $\cdot$ & 6000  & -455.554 $\pm$ 1.977 \\
        148.411 $\pm$ 0.89 & $\cdot$ & 6500  & -439.921 $\pm$ 5.392 \\
        152.823 $\pm$ 0.31 & $\cdot$ & 7000  & -418.264 $\pm$ 1.878 \\
        162.122 $\pm$ 0.23 & $\cdot$ & 8000  & -371.792 $\pm$ 2.281 \\
        171.820 $\pm$ 0.14 & $\cdot$ & 9000  & -322.728 $\pm$ 1.224 \\
    \end{tabular}
    \caption{Thermodynamic properties derived from spin-polarized two-phase simulations.}
    \label{two_phase_sp}
\end{table}

\newpage

\section*{Non-spin-polarized two-phase simulation results}

\begin{table}[h!]
    \centering
    \begin{tabular}{c c c c }
        \hline
        \hline
        Pressure (GPa) & Volume (\AA$^{3}$) & Temperature (K) & Internal Energy (eV) \\
        \hline
        45.806 $\pm$ 0.15 & 1162.83 & 5000  & -534.179 $\pm$ 0.815 \\
        49.914 $\pm$ 0.43 & $\cdot$ & 5500  & -515.134 $\pm$ 2.670 \\
        53.055 $\pm$ 0.18 & $\cdot$ & 6000  & -502.114 $\pm$ 1.729 \\
        56.172 $\pm$ 0.26 & $\cdot$ & 6250  & -482.124 $\pm$ 3.771 \\
        58.607 $\pm$ 0.32 & $\cdot$ & 6500  & -465.333 $\pm$ 2.843 \\
    \end{tabular}
    \caption{Thermodynamic properties derived from non-spin-polarized two-phase simulations.}
    \label{two_phase_ns}
\end{table}

\newpage

\section*{Homogeneous simulation results}

\begin{table}[h!]
    \centering
    \begin{tabular}{c c c c c}
        \hline
        \hline
        System & Volume (\AA$^{3}$/atom) & Enthalpy (eV/atom) & Entropy (N$k_B$) \\
        \hline
        Fe$_{64}$ & 11.33(3) & 0.539(6) & 18.29(11))\\
        Fe$_{32}$Mg$_{16}$O$_{16}$ & 10.93(5) & 0.592(18) & 17.29(10) \\
        Fe$_{16}$Mg$_{24}$O$_{24}$ &  10.52(5) & 0.288(17) & 16.21(9) \\
        Mg$_{32}$O$_{32}$ &  9.95(4) & -0.241(19) & 14.27(10) \\
        Fe$_{32}$Mg$_{48}$O$_{48}$  & 10.65(2) & 0.335(10) & 16.24(8) \\
    \end{tabular}
    \caption{Thermodynamic properties derived from spin-polarized homogeneous simulations at 60 GPa and 10,000 K.}
    \label{homogeneous_npt}
\end{table}

\newpage

\section*{Convergence of properties with respect to basis set size}

\begin{figure}[h!]
    \centering
    \includegraphics[width=0.95\textwidth]{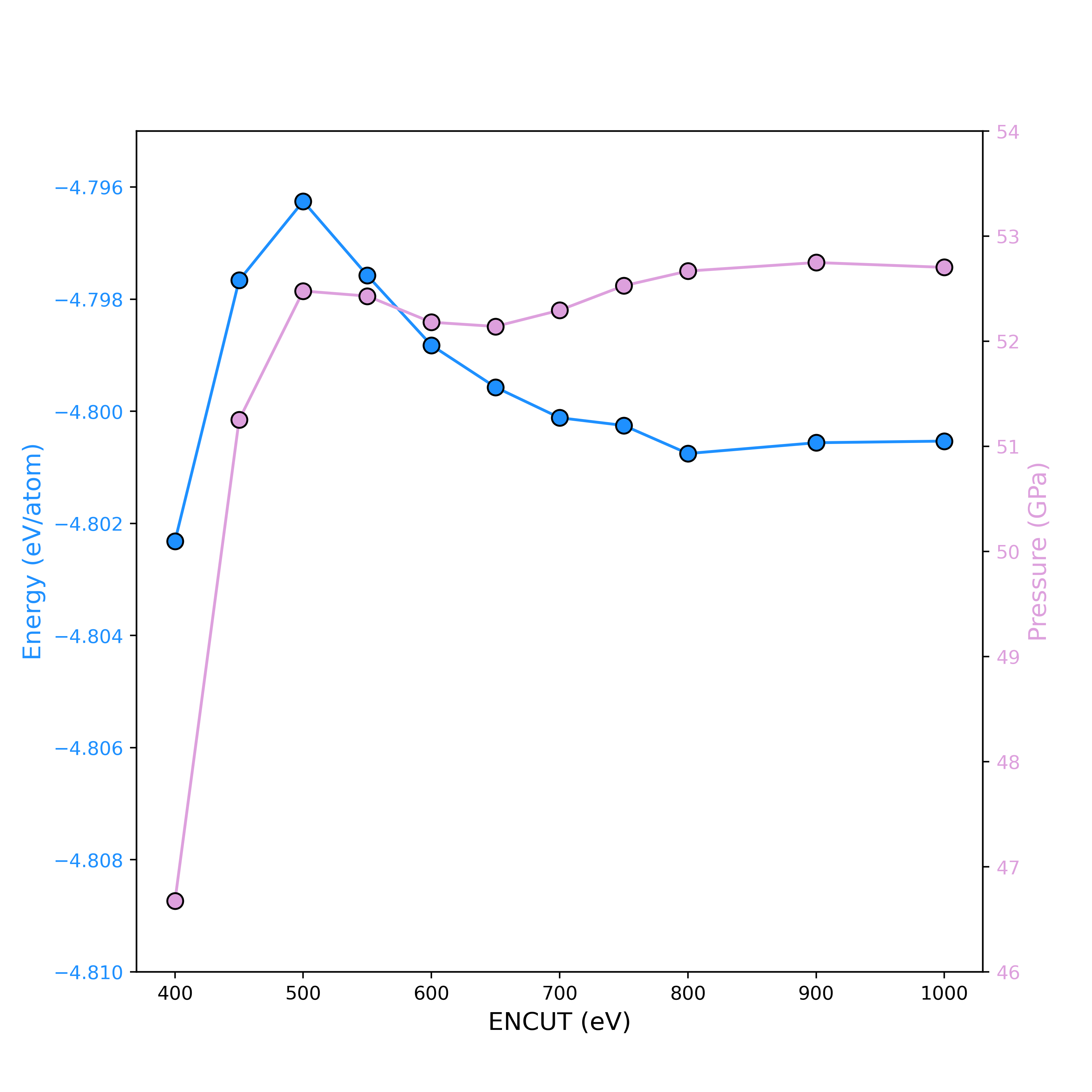}
    \caption{Internal Energy (blue, left-hand axis), and pressure (purple, right-hand axis) as a function of energy cutoff for an equilibrated snapshot of the two-phase simulation at 60 GPa and 6000 K.}
    \label{fig:enter-label}
\end{figure}

\newpage

\section*{Time evolution of phase compositions}

\begin{figure}[h!]
    \centering
    \includegraphics[width=0.95\textwidth]{willchand_evolve_6000.png}
    \caption{Time evolution of the compositions of the two coexisting fluids in the two-phase simulation at 6000 K, 61 GPa according to the method of Willard and Chandler (Eqs.~2-4 in the main text).  MgO-rich fluid (blue) and MgO-poor fluid (grey).  The compositions time-averaged over the equilibrated portion of the trajectory are $x_{MgO}=$ 93.99 \% $\pm$ 2.32 \% and 6.01 \% $\pm$ 2.32 \%, respectively.}
    \label{fig:enter-label}
\end{figure}

\newpage

\section*{Bond lifetimes}

\begin{figure}[h!]
    \centering
    \includegraphics[width=0.95\linewidth]{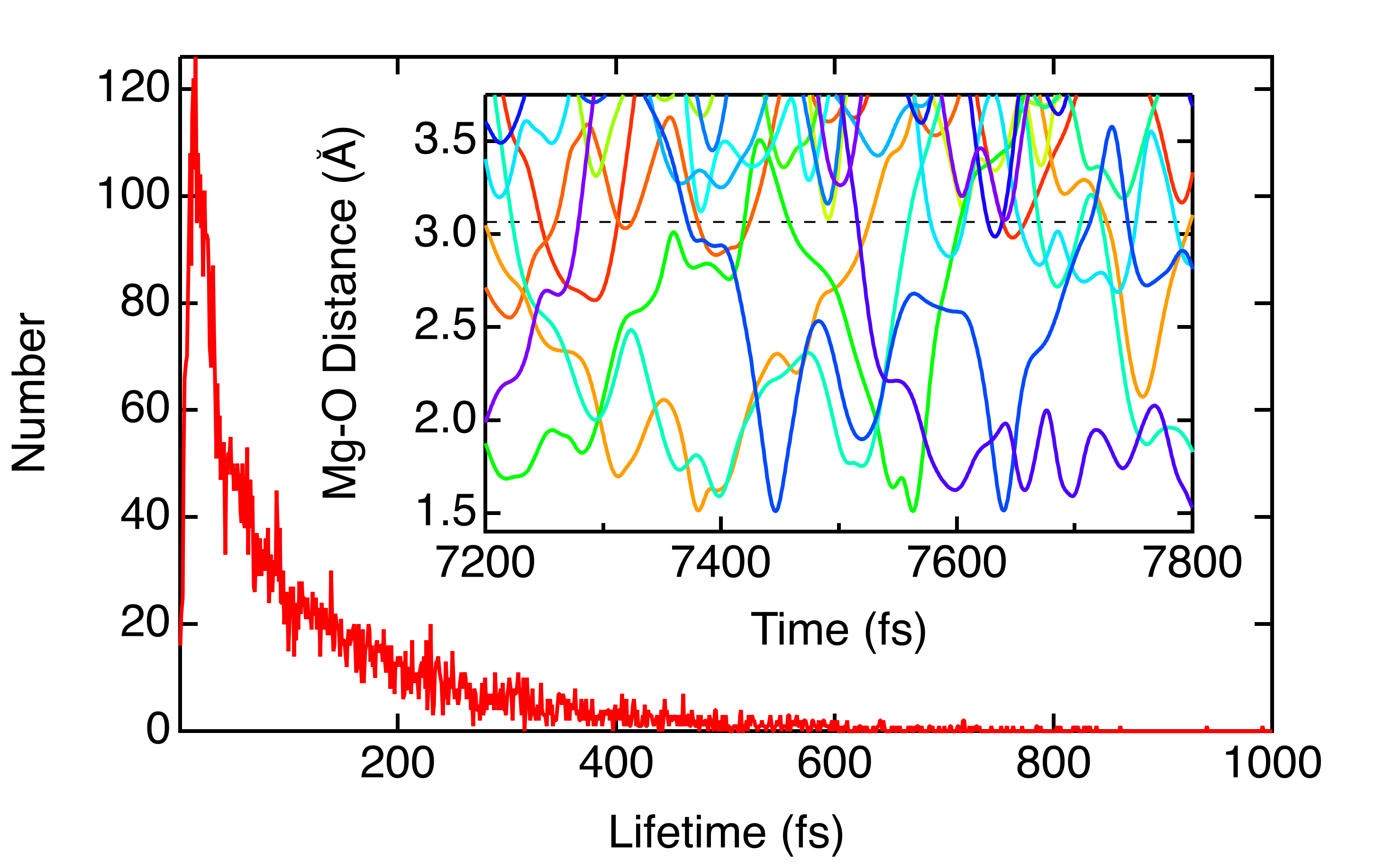}
    \caption{Main figure.  Histogram of Mg-O bond lifetimes at 10,000 K, 60 GPa, $x_{MgO}=0.6$.  Inset.  Distance of oxygen atoms from a single Mg atom over a portion of the trajectory (colored lines).  Each oxygen atom is given a unique color.  The black dashed line is the bond cutoff distance (3.08 \AA), which is taken to be the position of the first minimum in $g_{Mg-O}(r)$.  The mean Mg-O coordination number (4) is maintained via frequent exchanges of oxygen atoms within the coordination shell, for example the orange for the blue oxygen near 7520 fs.}
    \label{fig:enter-label}
\end{figure}









